\documentclass[prd,superscriptaddress,floatfix,notitlepage,nofootinbib,reprint]{revtex4-1} 

\usepackage{graphicx}  
\usepackage{dcolumn}   
\usepackage{bm}        
\usepackage{amsmath,amssymb}   
\usepackage{aas_macros}
\usepackage{multirow}
\usepackage{color}
\usepackage{verbatim}
\usepackage{url}
\usepackage{hyperref}

\newcommand{\mr}{\mathrm}

\newcommand{\bea}{\begin{eqnarray}}
\newcommand{\eea}{\end{eqnarray}}
\newcommand{\bmp}{\bm{\Psi}}
\newcommand{\bms}{\bm{s}}
\newcommand{\bmk}{\bm{k}}
\newcommand{\bmx}{\bm{x}}
\newcommand{\bmq}{\bm{q}}

\begin{document}

\widetext

\thispagestyle{empty}
\title{Nonlinear reconstruction}

\author{Hong-Ming Zhu}
\affiliation{Key Laboratory for Computational Astrophysics, National Astronomical Observatories, Chinese Academy of Sciences, 20A Datun Road, Beijing 100012, China}
\affiliation{University of Chinese Academy of Sciences, Beijing 100049, China}

\author{Yu Yu}
\affiliation{Department of Astronomy, Shanghai Jiao Tong University, 800 Dongchuan Road, Shanghai 200240, China}
\affiliation{Key Laboratory for Research in Galaxies and Cosmology, Shanghai Astronomical Observatory, Chinese Academy of Sciences, 80 Nandan Road, Shanghai 200030, China}

\author{Ue-Li Pen}
\affiliation{Canadian Institute for Theoretical Astrophysics, University of Toronto, 60 St. George Street, Toronto, Ontario M5S 3H8, Canada}
\affiliation{Dunlap Institute for Astronomy and Astrophysics, University of Toronto, 50 St. George Street, Toronto, Ontario M5S 3H4, Canada}
\affiliation{Canadian Institute for Advanced Research, CIFAR Program in Gravitation and Cosmology, Toronto, Ontario M5G 1Z8, Canada}
\affiliation{Perimeter Institute for Theoretical Physics, 31 Caroline Street North, Waterloo, Ontario N2L 2Y5, Canada}

\author{Xuelei Chen}
\affiliation{Key Laboratory for Computational Astrophysics, National Astronomical Observatories, Chinese Academy of Sciences, 20A Datun Road, Beijing 100012, China}
\affiliation{University of Chinese Academy of Sciences, Beijing 100049, China}
\affiliation{Center of High Energy Physics, Peking University, Beijing 100871, China}

\author{Hao-Ran Yu}
\affiliation{Kavli Institute for Astronomy and Astrophysics, Peking University, Beijing 100871, China}
\affiliation{Canadian Institute for Theoretical Astrophysics, University of Toronto, 60 St. George Street, Toronto, Ontario M5S 3H8, Canada}

\date{\today}

\begin{abstract}
We present a direct approach to nonparametrically reconstruct the linear
density field from an observed nonlinear map. 
We solve for the unique displacement potential consistent with the nonlinear 
density and positive definite coordinate transformation using a multigrid
algorithm. We show that we recover the linear initial conditions up to the 
nonlinear scale ($r_{\delta_r\delta_L}>0.5$ for $k\lesssim1\ h/\mathrm{Mpc}$) 
with minimal computational cost. 
This reconstruction approach generalizes the linear displacement theory to 
fully nonlinear fields, potentially substantially expanding the baryon acoustic
oscillations and redshift space distortions
information content of dense large scale structure surveys, including for 
example SDSS main sample and 21cm intensity mapping initiatives.
\end{abstract}

\maketitle


\section{Introduction}

The observation of cosmological large scale structure is
a cornerstone of modern cosmology. Ambitious surveys are mapping large swaths 
of the visible Universe (e.g. CHIME \cite{CHIME}, Tianlai \cite{Tianlai}, 
DESI \cite{desi}, PFS \cite{pfs}, and SDSS \cite{2016sdss}, etc). 
Precision measurements of baryon acoustic oscillations (BAO), redshift space 
distortions (RSD), and primordial non-Gaussianity, etc are continually
improving \cite{2017F,2017R,2016V,2017F2,2016S}.
The measured BAO scale can constrain the properties of dark energy and 
the growth rate measured from the RSD effect is crucial for tests of gravity.
However, the precision of the measurement is often limited by the strong 
non-Gaussianities of the dark matter and galaxy density fields on small scales, 
which prevent a simple mapping to the initial conditions that are predicted 
by cosmological theories.

The loss of the coherence to the initial conditions has been known as mode-mode
coupling, information saturation, etc. Some of the couplings are understood 
as arising from the coupling of large scale linear modes to smaller scale still
linear modes (e.g. cosmic tides \cite{tides1,tides2,tides3}, supersample 
covariance \cite{sc1,sc2,sc3}). These can be corrected by a linear mapping, 
also known as ``reconstruction'' \cite{2007bao}. 
The density fluctuations on mildly nonlinear scales can be roughly thought of
as the initial linear density fluctuations being translated by the bulk flows. 
The incoherent bulk flows destroy the coherence to the initial conditions. 
The density field reconstruction technique reverses the large scale bulk flows 
using the estimated displacement field \cite{2007bao}. However, the density 
field reconstruction methods based on the linear continuity equation only 
capture the effects of the large scale linear bulk flows instead of the full 
nonlinear bulk flows.

In this paper, we propose a new approach to reconstruct the linear density 
field through a nonlinear mapping, which removes most shift nonlinearities.
The reconstructed density field given by the displacement potential correlates 
with the initial linear field to $k\simeq1\ h/\mr{Mpc}$, about a factor of five
shorter length scale than observed in Eulerian space. 
This will substantially improve the measurements of BAO and RSD in the current
and future surveys.
The new reconstruction scheme offers an incisive tool for probing cosmology and particle physics.
We expect the new method to improve cosmological measurement techniques by 
orders of magnitude to answer many precise questions, e.g. neutrino masses, 
primordial non-Gaussianities, and modifications to gravity theories.

The paper is organized as follows. Section \ref{sec:algorithm} presents the
reconstruction algorithm. In Sec. \ref{sec:performance}, we apply nonlinear 
reconstruction to dark matter density field and show reconstruction results.
In Sec. \ref{sec:intepretations}, we present the physical interpretations for
the improved performance. In Sec. \ref{sec:applications}, we discuss future 
applications of the new reconstruction method.

\section{Reconstruction algorithm}
\label{sec:algorithm}

The basic idea is to build a bijective mapping
between the Eulerian coordinate system $\bm{x}$ and a new coordinate system 
$\bm{\xi}$, where the mass per volume element is constant. 
We define a coordinate transformation that is pure gradient,
\bea
\label{eq:mapping}
x^i=\xi^\mu\delta^i_\mu+\frac{\partial \phi}{\partial \xi^\nu}\delta^{i\nu},
\eea
where $\phi(\bm{\xi})$ is the {\it displacement potential} to be solved.
The new coordinate system is unique as long as we require the coordinate 
transformation defined above is positive definite, i.e., $\mr{det}(\partial x^i/\partial \xi^\alpha)>0$.
We call this new coordinate system {\it potential isobaric gauge/coordinates}.
It becomes analogous to ``synchronous gauge'' and ``Lagrangian coordinates''
before shell crossing, but allows a unique mapping even after shell crossing. 
Since the Jacobian of Eq. (\ref{eq:mapping}) is positive definite, we have
$\partial x^a/\partial \xi^a>0$ (no summation), from which it follows that 
each Eulerian coordinate is a monotonically increasing function of its 
corresponding potential isobaric coordinate and vice versa. 
This implies when we plot the Eulerian positions of the potential isobaric 
coordinates, the curvilinear grid lines will never overlap.

The unique displacement potential $\phi(\bm{\xi})$ consistent with the 
nonlinear density and positive definite coordinate transformation can be solved
using the moving mesh approach, which is originally introduced for the adaptive
particle-mesh $N$-body algorithm and the moving mesh hydrodynamics algorithm
\cite{1995ApJS..100..269P,1998ApJS..115...19P}.
The moving mesh approach evolves the coordinate system towards a state of 
constant mass per volume element, $\rho(\bm{\xi})d^3\xi=constant$. 
Since the shift from potential isobaric coordinates to Eulerian coordinates
can be large, the displacement potential must then be solved perturbatively.
We solve for a coordinate transformation $\Delta\phi^{(i)}$ at each step, 
where the shift $\nabla\Delta\phi^{(i)}$ is a small quantity, and then calculate
the density field in the new coordinate frame. 
The positive definiteness of the coordinate transformation is achieved through
smoothing and grid limiters \cite{1995ApJS..100..269P,1998ApJS..115...19P}.
We need to iterate this process for many times until the result converges and 
obtain the nonlinear bijective mapping from the Eulerian coordinate system to 
the potential isobaric gauge, 
$\phi=\Delta\phi^{(1)}+\Delta\phi^{(2)}+\Delta\phi^{(3)}+\cdots$, which results
from a continuous sequence of positive definite coordinate transformations.
The details of this calculation are given in Appendix \ref{appendix:A}.

We define the negative Laplacian of the reconstructed displacement potential
the {\it reconstructed density field},
\bea
\label{eq:rec}
\delta_r(\bm{\xi})\equiv-\nabla_{\bm{\xi}}\cdot\nabla_{\bm{\xi}}\phi(\bm{\xi})
=-\nabla_{\bm{\xi}}^2\phi(\bm{\xi}).
\eea
Note that the reconstructed density field is computed in the potential isobaric
gauge instead of the Eulerian coordinate system.

\section{Implementation and results}
\label{sec:performance}

To test the performance of the 
reconstruction algorithm, we run a $N$-body simulation with the ${\tt CUBEP^3M}$
code \cite{2013code}. The simulation involves $2048^3$ dark matter particles 
in a box of length $600\ \mr{Mpc}/h$ per side.
We use the snapshot at $z=0$ and generate the density field on a $512^3$ grid.
We solve for the displacement potential from the nonlinear density field and 
then have the reconstructed density field in the potential isobaric gauge.
The reconstruction code is mainly based on the {\tt CALDEFP} and {\tt RELAXING}
subroutines from the moving mesh hydrodynamics code \cite{1998ApJS..115...19P}.
The details of the numerical implementation are presented in Appendix \ref{appendix:A}.

Figure \ref{fig:map} shows a slice of the nonlinear dark matter density field.
We also plot the Eulerian position of each grid point of the potential 
isobaric gauge. The salient feature is the regularity of the grid. 
Even in projection, the grid never overlaps itself. This is guaranteed by 
appropriate smoothing and grid limiters \cite{1998ApJS..115...19P}.
The distribution of curvilinear grid points becomes denser in the higher 
density regions and sparser in the lower density regions; as a result 
the mass per curvilinear grid cell is approximately constant.

\begin{figure}[tbp]
\begin{center}
\includegraphics[width=0.48\textwidth]{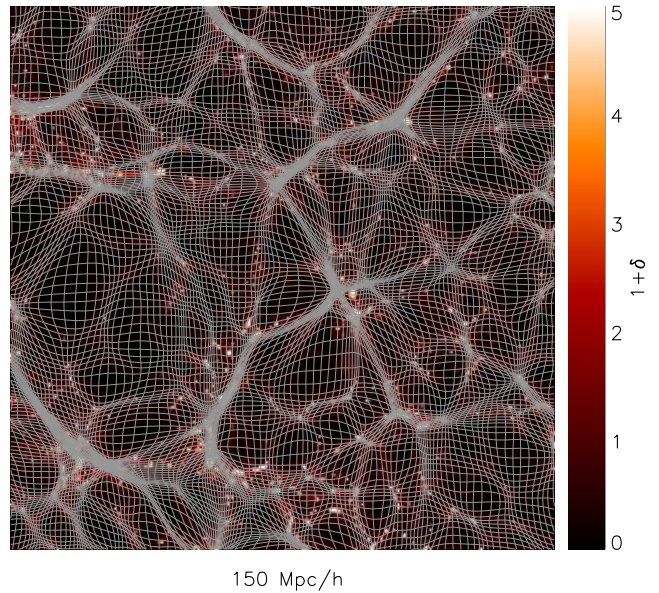}
\end{center}
\vspace{-0.7cm}
\caption{A slice of the nonlinear density field from the simulation. 
    The curvilinear grid shows the Eulerian coordinate of each grid point of
    the potential isobaric coordinate.}
\label{fig:map}
\end{figure}

\begin{figure}[tbp]
\begin{center}
\includegraphics[width=0.48\textwidth]{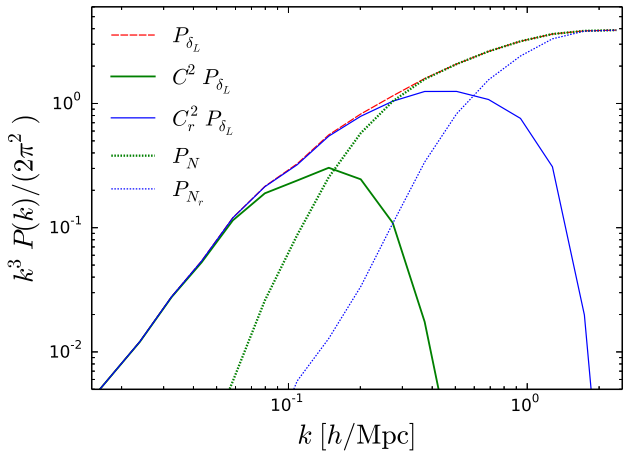}
\end{center}
\vspace{-0.7cm}
\caption{The linear power spectrum (dashed line), the linear signals for the 
nonlinear (thick solid line) and reconstructed (thin solid line) density fields,
and the mode-coupling terms for the nonlinear (thick dotted line) and reconstructed (thin dotted line) density fields. 
The curves are scaled such that the linear signal plus the mode-coupling term
equals the linear power. The linear signal is larger than the mode-coupling term
at $k\lesssim0.6\ h/\mr{Mpc}$ after reconstruction.}
\label{fig:ps}
\end{figure}

To directly quantify the information of the initial conditions in the density 
field, we calculate the propagator of the density field,
\bea
C(\bmk)=P_{\delta\delta_L}(\bmk)/P_{\delta_L}(\bmk),
\eea
where $\delta_L$ is the linear density field scaled to $z=0$ using the linear 
growth function. 
The matter power spectrum can be written as
\bea
P_\delta(\bmk)=C^2(\bmk)P_{\delta_L}(\bmk)+P_N(\bmk),
\eea
where $C^2(\bmk)P_{\delta_L}(\bmk)$ is the linear signal, which is the memory 
of the initial conditions, and $P_N(\bmk)$ is the power generated in the 
nonlinear evolution, often referred as the mode-coupling term
\cite{2006crocce,2008crocce,2008matsubara}.
Figure \ref{fig:ps} shows the linear signals and the mode-coupling terms 
for the nonlinear and reconstructed density fields. 
The linear signal is larger than the mode-coupling term at 
$k\lesssim0.6\ h/\mr{Mpc}$ after reconstruction, which
suggests that all BAO wiggles may be recovered from the present day density
field. Even the densest local Universe galaxy surveys such as the SDSS main
sample become Poisson noise dominated at this scale, opening up the potential 
of recovering cosmic information including BAO and potentially redshift-space
distortion down to the Poisson noise limit.

\begin{figure}[tbp]
\begin{center}
\includegraphics[width=0.48\textwidth]{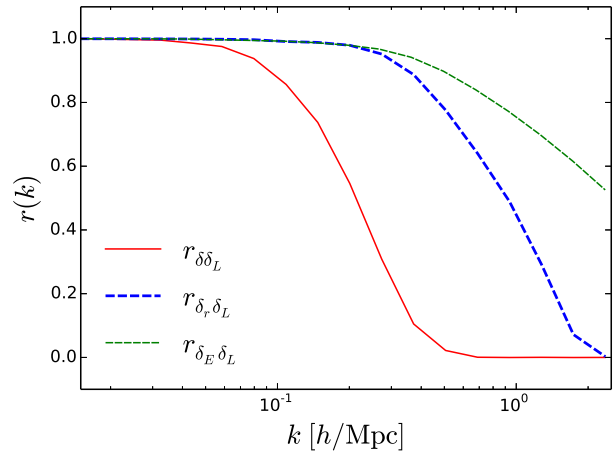}
\end{center}
\vspace{-0.7cm}
\caption{The cross-correlation coefficients with the initial conditions for the
nonlinear density field (solid line), the reconstructed density field 
(thick-dashed line), and the nonlinear displacement (thin-dashed line).}
\label{fig:cc}
\end{figure}

Reconstruction reduces the nonlinear damping of the linear power spectrum as
well as the mode-coupling term. To quantify the overall performance, we compute
the cross-correlation coefficients between the density field and the linear 
initial conditions,
\bea
r_{\delta\delta_L}(\bmk)=\frac{P_{\delta\delta_L}(\bmk)}
{\sqrt{P_\delta(\bmk)P_{\delta_L}(\bmk)}}=\frac{1}{\sqrt{1+\eta(\bmk)}},
\eea
where $\eta(\bmk)=P_N(\bmk)/(C^2(\bmk)P_{\delta_L}(\bmk))$ quantifies the 
relative amplitude of the linear signal to the mode-coupling term.
Figure \ref{fig:cc} shows the cross-correlation coefficients for the nonlinear 
and reconstructed density fields. We also plot the cross-correlation coefficient
of $\delta_E(\bmq)\equiv-\nabla\cdot\bmp(\bmq)$ with the linear density field,
where $\bmp(\bmq)$ is the nonlinear displacement from the simulation.
We note that the nonlinear displacement field correlates with the initial 
density field to even smaller Lagrangian scales.  This displacement field is
not actually observable, but presumably serves as a hard upper bound
on information that could plausibly be recovered from a nonlinear
density field, which is scrambled by shell crossing.  Several
improvements to this reconstruction approach may improve the
correlation further, for example using more grid cells or iteratively
improving the density field match (see Appendix \ref{appendix:B}).  
We leave this for future studies, since it would not likely improve 
the reconstruction from current galaxy surveys. 
The reconstruction performance on small scales will be limited by the nonlinear
galaxy bias and Poisson noise in galaxy surveys.

\begin{figure}[tbp]
\begin{center}
\includegraphics[width=0.48\textwidth]{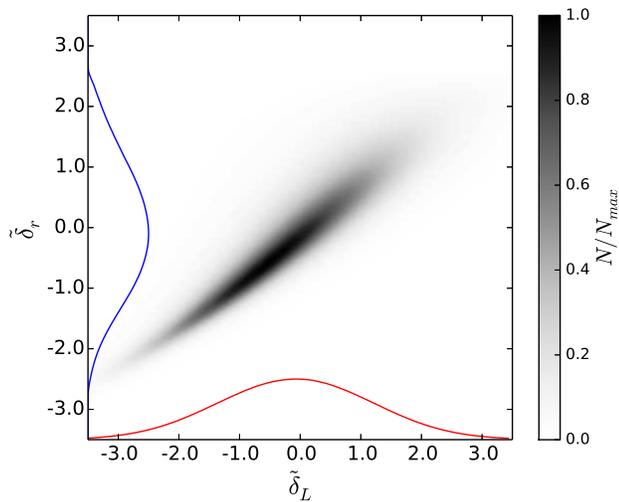}
\end{center}
\vspace{-0.7cm}
\caption{The joint PDF of the reconstructed and linear density field and the
marginal PDFs of the density fields. The tilde denotes the field has been 
Wiener filtered.}
\label{fig:pdf}
\end{figure}

In Fig. \ref{fig:pdf}, we show the joint probability distribution function (PDF)
of the reconstructed and linear density fields and the marginal PDFs of the 
density fields. 
Since the PDF depends on the grid scale, we apply the Wiener filter 
$W(\bmk)=C^2_r(\bmk)P_{\delta_L}(\bmk)/P_{\delta_r}(\bmk)$
to both fields to obtain the converged results.
The reconstructed densities are well correlated with the initial conditions and 
the PDF is also apparently much more Gaussian than the nonlinear density field. 
Therefore, the new reconstruction method is also expected to reduce the correlation between different power spectrum bins and increase the information content 
\cite{1999MNRAS.308.1179M,1999ApJ...527....1S,2005MNRAS.360L..82R}.
Note that all the reconstructed overdensities are smaller than 3 with the 
maximum value 2.693. The reconstructed density field is given by 
\bea
\delta_r(\bm{\xi})=-\nabla_{\bm{\xi}}\cdot(\bmx(\bm{\xi})-\bm{\xi})=
3-\nabla_{\bm{\xi}}\cdot\bmx(\bm{\xi}),
\eea
where $\nabla_{\bm{\xi}}\cdot\bmx(\bm{\xi})=\sum_a\partial x^a/\partial\xi^a$.
The compression limiter constrains $\partial x^a/\partial\xi^a\geq 0.1$, which
implies $\delta_r\leq2.7$ as we indeed observe in the reconstruction.
This confirms that the coordinate transformation given by the displacement 
potential defined in Eq. (\ref{eq:mapping}) is positive definite. 
In the 1D case, the maximum value of the reconstructed density field is smaller than 1 \cite{2016arXiv160907041Z}, since there is only one spatial dimension in the 1D cosmology \cite{2016matt}.

\section{Physical interpretations}
\label{sec:intepretations}

In the Lagrangian picture of structure 
formation, the Eulerian position $\bm{x}$ of each particle is given by the sum 
of its Lagrangian position $\bm{q}$ and the subsequent displacement 
$\bm{\Psi}(\bm{q})$:
\bea
\bm{x}(\bm{q})=\bm{q}+\bm{\Psi}(\bm{q}).
\eea
The density at the Eulerian position $\bm{x}$ is related to the displacement 
via the mass conservation $\rho(\bm{x})d^3x=\bar{\rho}d^3q$ or equivalently
$[1+\delta(\bm{x})]d^3x=d^3q$.
In the standard reconstruction approach \cite{2007bao}, the displacement field
is given by the linear continuity equation
\bea
\bms(\bmk)=\frac{i\bmk}{k^2}S(k)\delta(\bmk),
\eea
where $S(k)$ is a Gaussian window function which suppresses the small-scale 
nonlinearities.
The linear mapping defined by the estimated displacement field $\bms({\bmq})$ 
can transform the density field in Eulerian coordinates to the density field 
in Lagrangian coordinates through the mass conservation.
To compute the overdensities at the Eulerian positions $\bmx=\bmq+\bms(\bmq)$,
instead of assign particles to a curvilinear grid given by $\bms(\bmq)$, 
we can displace the particles by the negative displacement $-\bms(\bmq)$ and
then assign particles to a uniform grid to obtain the displaced density field 
$\delta_d(\bmx)$. To compute the Jacobian of the coordinate transformation, we 
can shift a set of uniformly distributed particles by the negative displacement
$-\bms(\bmq)$ and calculate the shifted density field $\delta_s(\bmx)$. 
The Jacobian of the mapping between Lagrangian and Eulerian coordinates is 
given by $[1+\delta_s(\bmx)]d^3x=d^3q$.
From the mass conservation $[1+\delta_r(\bmq)]d^3q=[1+\delta_d(\bmx)]d^3x$, 
we have the reconstructed density field
\bea
\delta_r(\bmq)=\bigg[1+\delta_d(\bmx)\bigg]\frac{d^3x}{d^3q}-1
\simeq\delta_d(\bmx)-\delta_s(\bmx),
\eea
where we assume that the shifted density field $\delta_s(\bmx)$ is small such 
that $[1+\delta_s(\bmx)]^{-1}\simeq1-\delta_s(\bmx)$.
Note that the reconstructed density field $\delta_r(\bmq)$ is not uniform in
the estimated Lagrangian coordinates since $\bms(\bmq)$ is the large-scale
linear displacement instead of the full nonlinear displacement $\bmp(\bmq)$.
In the new reconstruction approach, we solve the nonlinear mapping between the
Eulerian coordinate system and the potential isobaric gauge, where the mass per 
volume element is constant, and define the negative divergence of the estimated
displacement as the reconstructed density field.

In the nonlinear evolution, there are three sources of nonlinearities: bulk 
flows, shell crossing and structure formation \cite{2007ESW,2012TZ,2014Tassev}.
The decay of the propagator for the density field on the mildly nonlinear
scales is mainly due to the effects of the bulk motions.
The velocity power spectrum peaks at rather large scales, therefore the density
fluctuations on mildly nonlinear scales can be crudely thought of as the 
translated initial density fluctuations, where the translation is given by the  
displacement field \cite{2012TZ}.
The incoherent bulk flows destroy the memory of the initial conditions and cause
the decay of the propagator with the characteristic scale given by the root mean
square particle displacement. 
The standard BAO reconstruction approach uses the estimated displacement field
from the linear continuity equation to reduce the effects of the large-scale
bulk flows (the damping of the linear signal and the mode-coupling term) 
\cite{2009PWC,2009NWP,2016Seo}.
However, the new reconstruction scheme captures the full nonlinear displacement 
to the nonlinear (free-streaming) scale, where shell crossing occurs.
The dominant nonlinearities due to the mapping from Lagrangian coordinates to 
Eulerian coordinates are removed by nonlinear reconstruction except the 
nonlinearities induced by shell crossing.
The nonlinear contribution to the nonlinear displacement field also reduces the
correlation between the linear density and nonlinear displacement fields \cite{2016BSZ,2016yu}.
These nonlinearities arise from nonlinear clustering and therefore can not be 
removed by the nonlinear mapping from the new reconstruction approach.
The potential isobaric gauge avoids most of the shift nonlinearities induced by
the coordinate transformation except the inherent nonlinearities due to 
structure formation (the deviation of $r_{\delta_E\delta_L}$ from unity) and 
the residual shift nonlinearities due to shell crossing (the difference 
between $r_{\delta_r\delta_L}$ and $r_{\delta_E\delta_L}$).

\section{Applications}
\label{sec:applications}

The reconstructed density field correlates with the 
initial linear field to the nonlinear scale ($r_{\delta_r\delta_L}>0.5$ for 
$k\lesssim1\ h/\mr{Mpc}$) with the linear signal larger than the mode-coupling
term for $k\lesssim0.6\ h/\mr{Mpc}$.
We expect the reconstructed density field has a comparable fidelity as the 
linear density field for measuring the BAO scale, since the oscillations in the
linear power spectrum are also washed away on small scales.
The current BAO reconstruction displaces particles according to the displacement
field computed from the observed galaxy density field under some certain model
assumptions (the smoothing scale, galaxy bias, and growth rate etc).
The reconstruction result depends on the assumed fiducial model and must be
tested against different parameter choices. 
However, we directly solve the displacement potential from the observed density
field, which is a purely mathematical problem without any cosmological dynamics
involved. The implementation of the new reconstruction algorithm does not need
any model assumptions.

There do exist other methods can recover similar correlation with the linear 
initial conditions, e.g. the Hamiltonian Markov chain Monte Carlo method 
\cite{2014Wang}. However, the Hamiltonian sampling methods can only recover 
the phase correlation since they have assumed an initial linear power spectrum 
in the reconstruction. Thus, the Hamiltonian Markov chain Monte Carlo method 
cannot readily be applied to galaxy surveys to reconstruct the linear BAO 
signals, since the BAO peak location is already a model input.

The observed galaxy clustering pattern is anisotropic due to the RSD effect. 
The observed position of a galaxy is shifted from the true position by its 
peculiar velocity along the line of sight direction, which corresponds to a 
simple additive offset of the displacement. 
The reconstructed density field also includes the RSD effect. However, since
a lot of nonlinearities are removed by nonlinear reconstruction, both the
measurement and modeling of RSD will be improved significantly.
We have verified this, however, a detailed study is beyond the scope of this 
paper and will be presented in the future.

The current velocity reconstruction methods are based on the linear continuity 
equation. However, we generalize the linear displacement theory to fully 
nonlinear fields through nonlinear reconstruction. 
The new velocity reconstruction scheme based on the nonlinear fields is expected
to have better performance than the linear theory.
Moreover, we expect the new reconstruction method to improve the measurement 
techniques for the neutrino masses, primordial non-Gaussianities, modifications
to gravity theories, etc by orders of magnitude. 

\begin{acknowledgements}
We would like to thank Uros Seljak, Matias Zaldarriaga, Yin Li and Martin White
for valuable discussions.
We acknowledge the support of the Chinese MoST under Grant No. 2016YFE0100300, 
the NSFC under Grants No. 11633004, 
No. 11373030, No. 11403071 and No. 11773048, 
Institute for Advanced Study at Tsinghua University
and Natural Sciences and Engineering Research Council of Canada.
The simulation is performed on the BGQ supercomputer at the SciNet HPC 
Consortium. SciNet is funded by the Canada Foundation for Innovation under the 
auspices of Compute Canada, the Government of Ontario, Ontario Research 
Fund - Research Excellence, and the University of Toronto.
The Dunlap Institute is funded through an endowment established by the David Dunlap family and the University of Toronto.
Research at the Perimeter Institute is supported by the Government of Canada
through Industry Canada and by the Province of Ontario through the Ministry of
Research $\&$ Innovation.
\end{acknowledgements}
\appendix 

\section{Reconstruction algorithm}
\label{appendix:A}
In this Appendix, we present the details of the reconstruction
algorithm and its numerical implementation.

In Cartesian coordinates, the continuity equation of fluid dynamics, which
epresses the conservation of matter, is
\bea
\label{eq:continuity}
\frac{\partial\rho}{\partial t}+\frac{\partial\rho v^i}{\partial x^i}=0,
\eea
where $\rho$ is the fluid density, $\bm{v}$ is the fluid velocity, and
$\rho\bm{v}$ is the mass flux density.
The total mass of fluid flowing out of a volume element $d^3x$ in unit time
$\nabla\cdot(\rho\bm{v})d^3x$ is the decrease per unit time in the mass of
fluid in this volume element $(-\partial\rho/\partial t)d^3x$.

We apply a general time-dependent curvilinear coordinate transformation
$\bm{x}=\bm{x}(\bm{\xi},t)$ to the continuity equation and obtain
\bea
\label{eq:continuityxi}
\frac{\partial\sqrt{g}\rho}{\partial t}+\frac{\partial}{\partial\xi^\alpha}
\bigg[\sqrt{g}\rho e^\alpha_i\bigg(v^i-\dot{x}^i\bigg)\bigg]=0,
\eea
where $e^\alpha_i$ is the matrix inverse of the triad $e^i_\alpha=\partial x^i/
\partial\xi^\alpha$, and $\sqrt{g}=\mr{det}(e^i_\alpha)$ is the volume element.
This is the continuity equation in the time-dependent curvilinear coordinate
system. However, it also describes the change of the mass per volume element
under the time-dependent coordinate transformation $\dot{\bm{x}}$ if the fluid
velocity is zero. We can use this equation to evolve the curvilinear coordinate
system toward a state of constant mass per volume element across the Universe.

Since we can only observe the density field from galaxy surveys, this allows
to determine only the scalar part of the coordinate transformation due to
the limited degrees of freedom.
We define a coordinate transformation that is a pure gradient
$x^i=\xi^\mu\delta^i_\mu+({\partial \phi}/{\partial \xi^\nu})\delta^{i\nu}$,
and set the velocity in Eq. (\ref{eq:continuityxi}) to zero.
This results in a linear elliptic evolution equation for the displacement
potential $\phi$:
\bea
\label{eq:elliptic_diff}
\partial_\mu(\rho\sqrt{g}e^\mu_i\delta^{i\nu}\partial_\nu\dot{\phi})
=\partial_t(\sqrt{g}\rho),
\eea
where $\dot{\phi}$ is the differential coordinate transformation and
$\partial_t(\sqrt{g}\rho)$ is the increase per unit time in the mass per unit
curvilinear coordinate volume.
We use the deviation density $\Sigma=\bar{\rho}-\rho\sqrt{g}$ as the desired
change of the mass per volume element,
\bea
\label{eq:elliptic}
\partial_\mu(\rho\sqrt{g}e^\mu_i\delta^{i\nu}\partial_\nu\dot{\phi})
=S(\Sigma+C+E),
\eea
where $S$ is the smoothing operator, $C$ is the compression limiter, and
$E$ is the expansion limiter \cite{1995ApJS..100..269P,1998ApJS..115...19P}.
We define the compression limiter $C$ and the expansion limiter $E$ as
\begin{align}
C(\phi)&\equiv4\bigg[\frac{\xi_m}{\lambda_0}-H\bigg(\frac{\xi_m}{\lambda_0}
-1\bigg)\bigg]^2,\\
E(\phi,\Sigma)&\equiv-2H(\sqrt{g}-v_m)|\Sigma|,
\end{align}
where $H$ is the Heaviside function, $\xi_m\approx1/10$ is the maximal
compression factor, $\lambda_0$ is the minimum eigenvalue of the triad
$e^i_\mu$. We choose a typical expansion volume limit $v_m=10$.
The smoothing operator $S$ is simplest to implement by smoothing over the
nearest neighbors in curvilinear coordinates.

We approximate $\bm{x}$ as $\bm{\xi}$ in Eq. (\ref{eq:elliptic}) and solve for
the differential coordinate transformation $\dot{\phi}$ using the multigrid
algorithm as described in Ref. \cite{1995ApJS..100..269P}.
We then calculate the exact change of the mass per volume element $\Delta\rho$
using Eq. (\ref{eq:elliptic_diff}),
\bea
\partial_\mu(\rho\sqrt{g}e^\mu_i\delta^{i\nu}\partial_\nu\Delta\phi)
=\Delta\rho,
\eea
where $\Delta\phi=S\dot{\phi}$, and obtain the density field in the new
coordinate frame, $\rho'=\rho+\Delta\rho$.
We iterate this process for many times until the mass per volume element is
approximately constant and obtain the displacement potential,
$\phi=\Delta\phi^{(1)}+\Delta\phi^{(2)}+\Delta\phi^{(3)}+\cdots$,
where $\Delta\phi^{(i)}$ is the solution from the $i$th iteration.

\section{Convergence tests}
\label{appendix:B}

\begin{figure}[tbp]
\begin{center}
\includegraphics[width=0.48\textwidth]{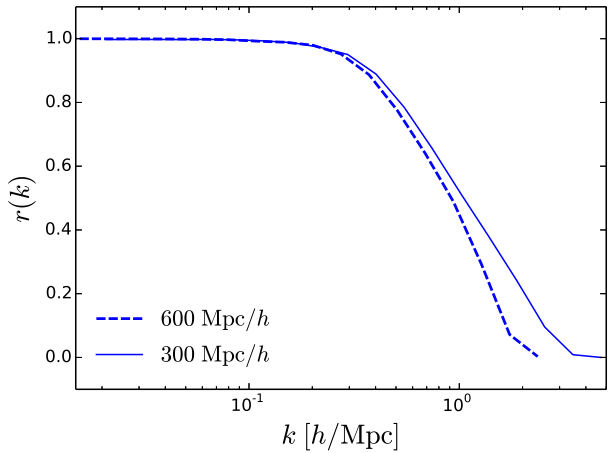}
\end{center}
\vspace{-0.7cm}
\caption{The cross-correlation coefficients between the reconstructed density
field and the linear initial conditions for the two simulations with box
size $600\ \mr{Mpc}/h$ (dashed line) and box size $300\ \mr{Mpc}/h$
(solid line).}
\label{fig:box}
\end{figure}

To check the convergence of reconstruction, we run a simulation with $1024^3$
dark matter particles in a box of side length $300\ \mr{Mpc}/h$.
Due to the sheer computational cost of multigrid calculation with a $1024^3$
grid, we instead apply reconstruction to the density field on a $512^3$
grid from this small box size simulation.
Figure \ref{fig:box} shows the cross-correlation coefficients between the reconstructed density field and the linear initial conditions for the two simulations.
We note that using more grid cells can further improve the correlation slightly.
However, it would not likely improve the reconstruction from current galaxy
surveys because of the nonlinear galaxy bias and Poisson noise on these scales.

\begin{figure}[tbp]
\begin{center}
\includegraphics[width=0.45\textwidth]{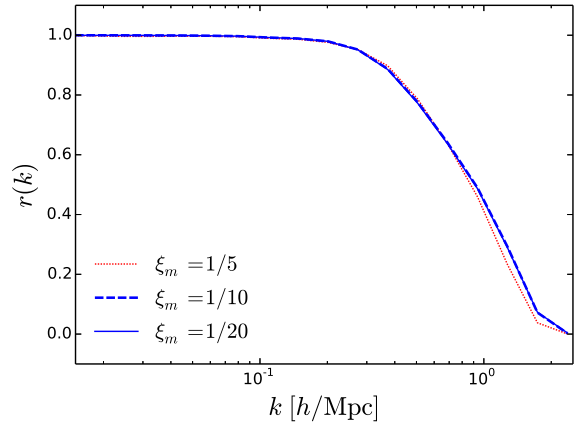}
\end{center}
\vspace{-0.75cm}
\caption{The cross-correlation coefficients between the reconstructed density
field and the linear initial conditions for the maximal compression factor
$\xi_m=(1/5,1/10,1/20)$ and the expansion volume limit $v_m=10$. }
\label{fig:cm}
\end{figure}

\begin{figure}[tbp]
\begin{center}
\includegraphics[width=0.45\textwidth]{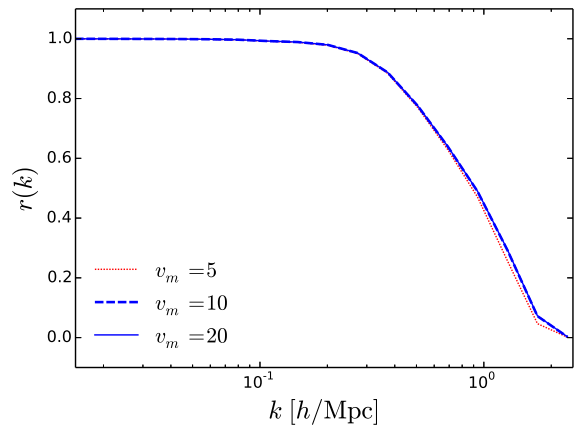}
\end{center}
\vspace{-0.75cm}
\caption{The cross-correlation coefficients between the reconstructed density
field and the linear initial conditions for the expansion volume limit
$v_m=(5,10,20)$ and the maximal compression factor $\xi_m=1/10$.
}
\label{fig:vm}
\end{figure}

To study how the reconstruction depends on the maximal compression factor
$\xi_m$ and expansion volume limit $v_m$, we perform reconstruction with
different $\xi_m$ and $v_m$.
We first set $\xi_m=(1/5,1/10,1/20)$ and keep $v_m=10$ and apply
reconstruction to the nonlinear density field.
Figure \ref{fig:cm} shows the cross-correlation coefficients of the
reconstructed density field with the linear initial conditions for different
values of the maximal compression factor.
We note that the reconstruction result converges for $\xi_m\lesssim1/10$.
To prevent excessive compression and the associated computational cost, we
choose $\xi_m\approx1/10$ in the reconstruction.
Next we set $v_m=(5,10,20)$ and keep $\xi_m=1/10$ and apply reconstruction to
the nonlinear density field.
Figure \ref{fig:vm} shows the cross-correlation coefficients of the
reconstructed density field with the linear initial conditions for different
values of the expansion volume limit.
We note that the reconstruction result converges for $v_m\gtrsim10$, so we
choose $v_m=10$.
We expect that $\xi_m=1/10$ and $v_m=10$ will the optimal choice for most cases
of reconstruction in the current galaxy surveys.

\bibliographystyle{apsrev}

\bibliography{3d}

\label{lastpage}
\end{document}